\documentclass[apjl]{emulateapj}

\usepackage{graphicx}
\usepackage{t1enc}
\usepackage[varg]{txfonts}
\usepackage{epsfig}
\usepackage{rotating}
\usepackage{natbib}
\usepackage{color}

\definecolor{RED}{rgb}{1.0,0.0,0.0}

\shorttitle{Warm HCN in the planet formation zone of GV~Tau~N}
\shortauthors{Fuente et al.}

\begin{document}

\title{Warm HCN in the planet formation zone of GV~Tau~N
\footnote{Based on observations carried out with the IRAM Plateau de Bure Interferometer. 
IRAM is supported by INSU/CNRS (France), MPG (Germany) and IGN (Spain).}
}

\author{Asunci\'on Fuente\altaffilmark{1},
Jos\'e Cernicharo\altaffilmark{2},
Marcelino Ag\'undez\altaffilmark{2,3,4}
}
\altaffiltext{1}{Observatorio Astron\'omico Nacional (OAN,IGN), Apdo 112, E-28803 Alcal\'a de Henares, Spain}
\email{a.fuente@oan.es}
\altaffiltext{2}{Centro de Astrobiolog\'{\i}a (CSIC/INTA), Laboratory of Molecular Astrophysics, Ctra. Ajalvir km. 4, E-28850, Torrej\'on de Ardoz, Spain}
\altaffiltext{3}{Univ. Bordeaux, LAB, UMR 5804, F-33270, Floirac, France}
\altaffiltext{4}{CNRS, LAB, UMR 5804, F-33270, Floirac, France}

\begin{abstract}
The Plateau de Bure Interferometer has been used to map the continuum emission
at 3.4~mm and 1.1~mm together with the J=1$\rightarrow$0 and J=3$\rightarrow$2 lines of
HCN and HCO$^+$ towards the binary star GV~Tau. The 3.4~mm observations did not resolve the binary components and
the HCN J=1$\rightarrow$0 and HCO$^+$ J=1$\rightarrow$0 line emissions trace the circumbinary disk and the flattened envelope. 
However, the 1.1~mm observations resolved the individual disks of GV~Tau~N and GV~Tau~S and allowed us to study their chemistry. 
We detected the HCN 3$\rightarrow$2 line only towards the individual disk of GV~Tau~N, and the emission of the HCO$^+$ 3$\rightarrow$2 line 
towards GV~Tau~S. Simple calculations indicate that 
the 3$\rightarrow$2 line of HCN is formed in the inner R$<$12~AU of the disk around GV~Tau~N where 
the HCN/HCO$^+$ abundance ratio is $>$300. On the contrary, 
this ratio is $<$1.6 in the disk around GV~Tau~S. The high HCN abundance 
measured in GV~Tau~N is well explained by photo-chemical processes 
in the warm ($>$400~K) and dense (n$>$10$^7$~cm$^{-3}$) disk surface.
\end{abstract}


\section{Introduction \label{sint}}
A fraction of the gas and dust in
protoplanetary disks will end up in planets and may constitute
the basis to form prebiotic species.
A large effort has been done aiming to detect
the warm gas in the planet formation zone
using Spitzer and NIR ground-based facilities. In a pionering work, Lahuis et al. (2006) detected strong HCN and C$_2$H$_2$ absorption features toward 
one source, IRS~46, from a sample of more than
100 Class I and II sources located in nearby star-forming regions. Later, Gibb et al. (2007, 2008) detected the HCN and C$_2$H$_2$ absorption lines in GV Tau.
Carr \& Najita (2011) detected the rotational transitions of H$_2$O and OH and the rovibrational bands of simple
organic molecules (CO$_2$, HCN, C$_2$H$_2$) in  11 classical T Tauri stars showing that these molecules are not uncommon in the 
inner region of the T Tauri disks. 
Thus far, only NIR spectroscopy has provided information about the chemical composition of the gas in the inner disk region.
Interferometric mm and submm observations are key to derive 
the kinematics and physical conditions  of the gas,
as well as the molecular abundance radial profiles that NIR studies cannot provide. 

GV~Tau (Haro 6-10) is a T Tauri binary system embedded in the L1524 molecular (d=140~pc)
cloud. It is one of a small number of young binaries for which the primary (GV~Tau~S)
is optically visible and the 
companion (GV~Tau~N), 
located 1.2$"$ to the North,
is strongly embedded. It is associated with a parsec-scale Herbig-Haro flow which extends to
1.6~pc to the north at a position angle of about 222$^\circ$ and $\sim$1~pc to the south 
(Devine et al. 1999; see also Movsessian \& Magakian 1999).
This region has been extensively studied in the NIR and all the studies pointed to the existence of a complex system composed of
two circumstellar disks associated with GV Tau N and GV Tau S respectively that are themselves surrounded by a circumbinary disk
and/or flattened nebula (M\'enard et al. 1993, Koresko et al. 1999, Leinert et al. 2001). Doppmann et al. (2008) and more 
recently Wilking et al. (2012) proposed 
that GV~Tau~N and possibly GV~Tau~S could be binary systems themselves. The two individual disks GV~Tau~N and GV~Tau~S are 
misaligned and present different inclination angles. The GV~Tau~N disk has an edge-on geometry, while the GV~Tau~S disk is closer to a 
face-on orientation which allows the stellar radiation to escape forming a visible nebula ( Roccatagliata et al. 2011). 
Only the GV~Tau~N disk has been detected in the H$_2$ 2.12~$\mu$m ro-vibrational line (Herbst et al. 1995). The
HCN and C$_2$H$_2$ absorption ro-vibrational lines are detected towards GV~Tau~N (Gibb et al. 2007, 2008) 
which proves the existence of a rich organic chemistry in the inner disk of this star. The non-detection of these
lines towards GV~Tau~S did not allow, however, to conclude about its disk chemistry since
it could be due to the different disk inclination.

We present 3.4~mm and 1.1~mm interferometric images of GV~Tau using the IRAM
Plateau de Bure Interferometer (PdBI).
The highest spatial resolution observations at 1.1~mm allowed us to resolve
the GV~Tau~N and GV~Tau~S individual disks
and have a first glance of the chemical differences between them.

\begin{figure*}
\includegraphics{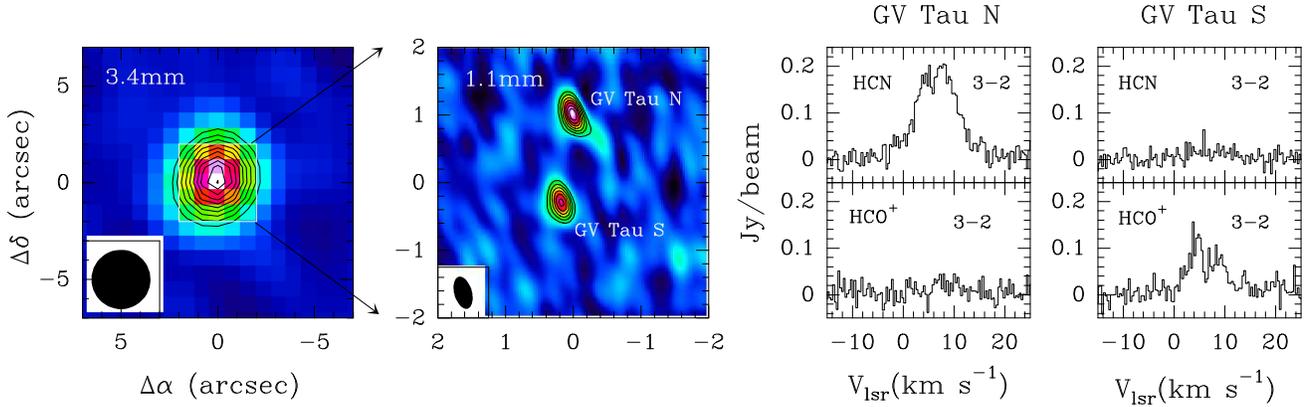}
\vspace{5.75cm}
      \caption{{\it Right panel:} Interferometric image of the continuum emission at 3.4~mm. Contour Levels are 3~mJy/beam to 13~mJy/beam by 1~mJy/beam.The image
is centered at RA(J2000)=04:29:23.73 Dec(J2000)=24:33:00.30. {\it Central panel}  Interferometric image of the continuum emission at 1.1~mm. 
Contour Levels are 10~mJy/beam to 50~mJy/beam by 5~mJy/beam.{\it Left panel} Interferometric spectra of the HCN 3$\rightarrow$2 and HCO$^+$ 3$\rightarrow$2 lines 
towards GV~Tau~N and GV~Tau~S.}
         \label{Fig 1}
 \end{figure*}

\begin{figure}
\includegraphics{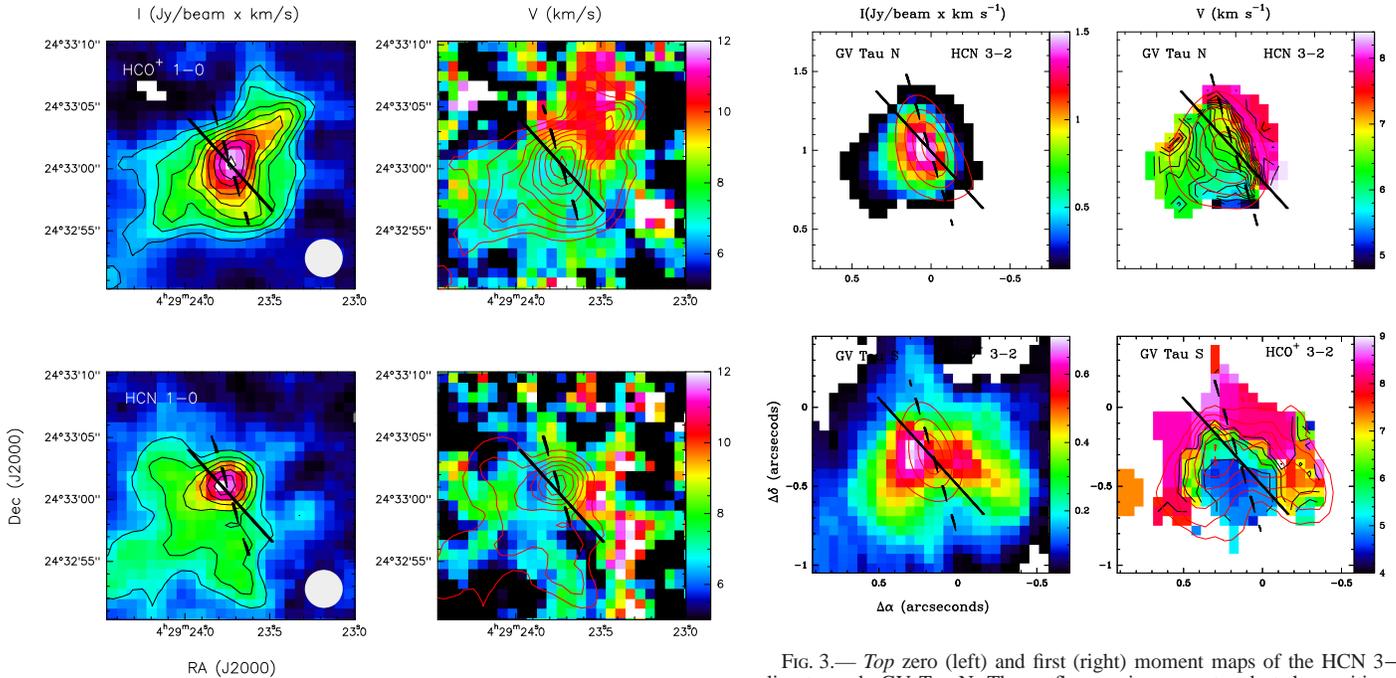}
\vspace{9.75cm}
      \caption{{\it Top} zero (left) and first (right) moment maps of the HCO$^+$ 1$\rightarrow$0 line towards GV~Tau.
Contour levels are 0.1 to 0.5 by 0.05~Jy/beam$\times$km~s$^{-1}$.
The directions of the large scale outflow defined by the giant Herbig Haro flow (PA=222$^\circ$; Devine et al. 1999) and of the
Herbig-Haro jet (PA=195$^\circ$; Movsessian \& Magakian 1999)
are indicated by two black lines through the phase center. {\it Bottom} the same for the HCN 1$\rightarrow$0 line.   }
         \label{Fig 2}
 \end{figure}

\section{Observations \label{sobs}}
The HCN 1$\rightarrow$0 and HCO$^+$ 1$\rightarrow$0 lines were observed using the PdBI in its CD configuration during 
2009 October-November. This configuration provided an angular resolution of  3.07$"$$\times$3.05$"$ PA 17$^\circ$  ($\sim$432~AU$\times$429~AU at the distance of Taurus). 
During the observations two 40 MHz bandwidth correlator units were placed at the frequencies of the HCN 1$\rightarrow$0 (88631.85~MHz) and HCO$^+$ 1$\rightarrow$0  (89188.52~MHz) lines 
providing a spectral resolution of 78 kHz. These lines were also observed with the 320~MHz units which provided a spectral resolution of $\sim$2.5~MHz. 
Other two 320~MHz units were placed in a 
frequency range free of lines to measure the continuum and subtract it from the spectral line maps.

In 2011 February, we performed sub-arcsecond imaging of this region in the HCN 3$\rightarrow$2 
and HCO$^+$ 3$\rightarrow$2 lines using the PdBI 
in its A configuration,
which provides a beam of $\sim$0.49$"$$\times$0.26$"$ PA 16$^\circ$ ($\sim$69~AU$\times$36~AU). Two 40~MHz correlator units 
were placed at the frequencies of 
the HCN 3$\rightarrow$2 (265886.18~MHz) and HCO$^+$ 3$\rightarrow$2 (267557.526~MHz) lines. We used the  correlator WIDEX to cover the 
4~GHz bandwidth of the receivers (264.58 - 268.18 GHz).
Only HCO$^+$ 3$\rightarrow$2 and HCN 3$\rightarrow$2 were detected 
down to an rms of 6~mJy/beam with a spectral resolution of 2~MHz. 
The channels without line emission in WIDEX were used to create the continuum 
1.1~mm map which was subtracted from the spectral maps.
The resulting 3.4~mm and 1.1~mm continuum images are shown in Fig.~1. The spectral maps are shown in Fig.~2 and Fig.~3. The phase center in our observations 
was RA(J2000)=04:29:23.73 Dec(J2000)=24:33:00.30, in between the two stars.

\begin{figure}
\includegraphics{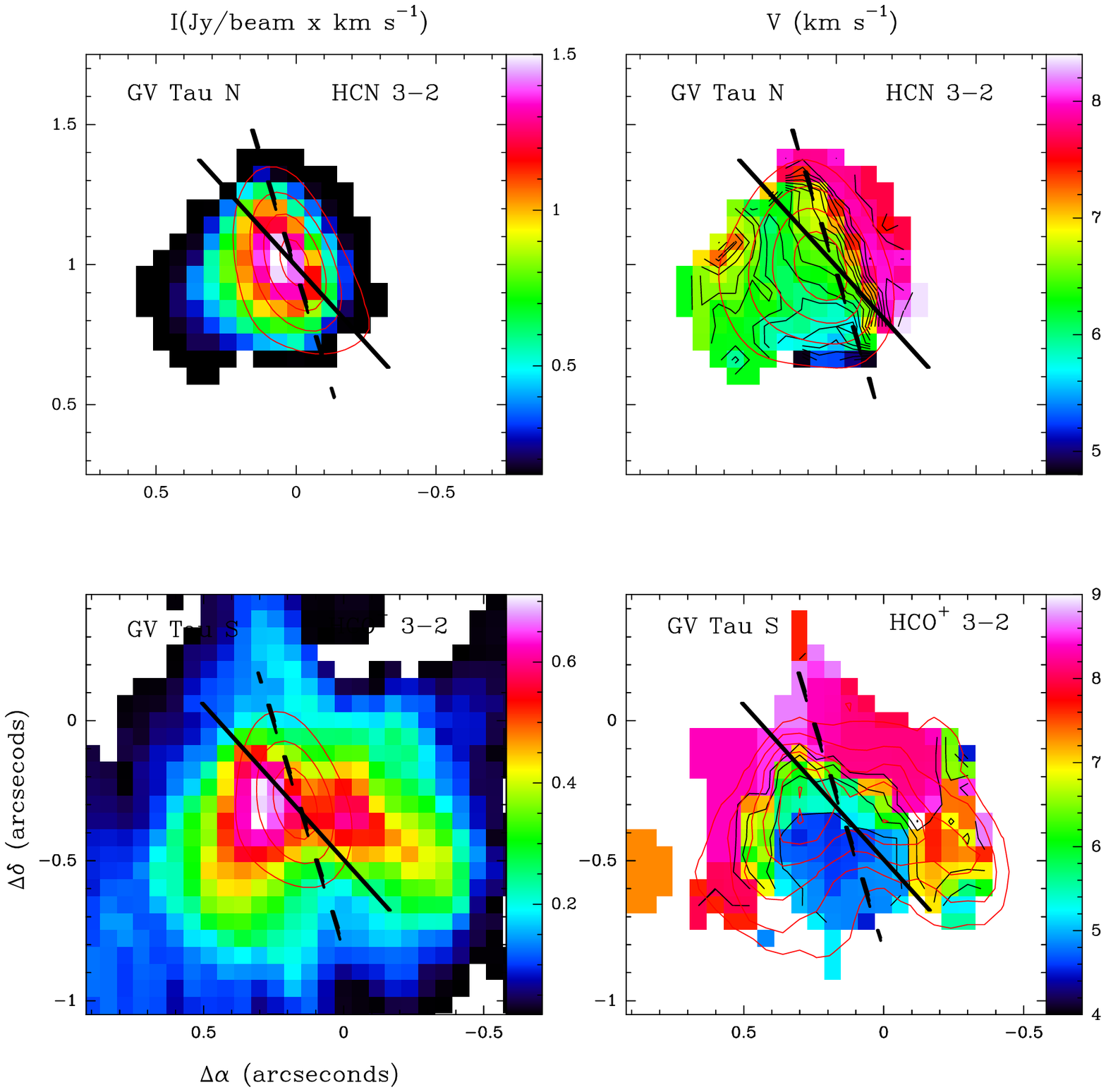}
\vspace{9.0cm}
      \caption{{\it Top} zero (left) and first (right) moment maps of the HCN 3$\rightarrow$2 line towards GV~Tau~N. The outflows axis are 
centered at the position of GV~Tau~N (04:29:23.731, 24:33:01.30).
In the left panel, the red contours correspond to the continuum 1.1~mm image and contour levels are
0.01 to 0.04 by 0.01~Jy/beam. In the right panel, red contours corespond to the integrated intensity of the HCN 3$\rightarrow$2 
line and contour levels are 0.150 to 1.5 by 0.150~Jy/beam$\times$km~s$^{-1}$. 
{\it Bottom} the same for the HCO$^+$ 3$\rightarrow$2 line. The outflow axes are centered at the position of GV~Tau~S (04:29:23.743, 24:32:59.99).
Contours in the left panel are 0.1 to 0.7 by 0.1~Jy/beam$\times$km~s$^{-1}$.
 }
         \label{Fig 3}
 \end{figure}

\section{Results \label{sres}}
\subsection{Continuum}
Our 3.4~mm image does not resolve the two components of the binary. An elliptical Gaussian was fit to the uv table and the parameters 
are shown in Table 1. The size of the emission, $\sim$280 AU in diameter, is slightly smaller than
the size of the NIR nebula observed by M\'enard et al. (1993). The 3.4~mm continuum emission peaks in the middle 
of the two stars, suggesting a similar contribution to the total flux from the two disks. The total flux, 16.1$\pm$0.4 mJy  is consistent 
with that measured by Guilloteau et al (2011).

Our 1.1~mm observations clearly resolve the two components which present similar intensity. Gaussian fits to the
uv tables were done and the parameters are shown in Table 1. Because of our elongated beam, the disks are resolved in only one direction.  Our fit gives
a radius of $\sim$0.07$"$ ($\sim$10~AU) for the emission of the two disks. Guilloteau et al. (2011) observed this target using the PdBI with a larger beam of
 0.89$"$$\times$0.56$"$ and derived a size twice larger than ours. After modeling the emission, they 
 concluded that the two disks are, very likely, optically thick and the size was overestimated because of the atmospheric seeing. This could be
 a reason for our disagreement. Another possibility is that we have filtered out part of the disk emission. The filtering effect would be more severe 
towards GV Tau S disk because of the face-on orientation. With the sizes derived from our observations, we estimate a 
brightness temperature of $>$69~K
for the N and  $>$56~K for the S components respectively (we give a lower limit because the emission is essentially unresolved in one direction), 
consistent with the emission arising from the R$<$10~AU
region of a T Tauri disk (Pinte et al. 2006).

\begin{table*}
{\scriptsize
\caption{Gaussian fits to the continuum and spectral data}
\label{tab_observations}
\begin{center}
\begin{tabular}{llllccccc} \hline 
 \multicolumn{3}{l}{Source}  &  \multicolumn{1}{c}{$\lambda$}  &  \multicolumn{1}{c}{HPBW($"$)} & \multicolumn{1}{c}{Flux (mJy)} &
\multicolumn{1}{c}{Major ($"$)}  & \multicolumn{1}{c}{Minor($"$)}  &  \multicolumn{1}{c}{PA}     \\  \hline
GV Tau N+S & 04:29:23.733 &  24:33:00.56  &  3.4mm$^1$  & 3.07$"$$\times$3.05$"$ & 16.1 (0.4)    &  2.02 ( 0.08)  &  1.06 ( 0.13)  & -14 (3) \\ \hline
GV Tau N   & 04:29:23.731 &  24:33:01.30  &  1.1mm$^1$  & 0.49$"$$\times$0.26$"$ &  49.8 ( 1.1)  &  0.14 ( 0.03)  &    0.03 (0.04)  &   52 (1) \\           
           &              &               &  1.3mm$^2$  & 0.89$"$$\times$0.56$"$ &  43.8 (3.1)  &  0.24  (0.11)  & 0.09 (0.06)   &  53 (18) \\  
           &              &               &  HCN~3$\rightarrow$2$^1$ &  0.50$"$$\times$0.25$"$  & 3.64(0.21)$^3$  &  0.32(0.04)  &  0.25(0.05)  &  -41(20) \\ \hline
GV Tau S   & 04:29:23.743 &  24:32:59.99  &  1.1mm$^1$  & 0.49$"$$\times$0.26$"$  & 38.0 ( 1.1)  &  0.12 ( 0.02)  & 0.03 (0.05)  &  -34 (1)  \\
           &              &               &  1.3mm $^2$ & 0.89$"$$\times$0.56$"$  & 46.7 (3.2)   &  0.37 (0.05)  & 0.11 (0.07)  &    -2(8)   \\   
           &              &               &  HCO$^+$~3$\rightarrow$2$^1$ & 0.49$"$$\times$0.28$"$ &  9.24(1.1)$^3$  &  1.26(0.06) &  0.73(0.09)  &  -74(5) \\
\hline
\end{tabular}
\end{center}
$^1$ This work; $^2$ Guilloteau et al. (2011); $^3$ Total flux in Jy$\times$km~s$^{-1}$.
}
\end{table*}

\subsection{Spectral Observations}

We have detected intense emission of the HCN~1$\rightarrow$0 and HCO$^+$~1$\rightarrow$0 lines towards GV~Tau. 
The integrated emission of the HCO$^+$ 1$\rightarrow$0 line is composed of a compact source centered at the phase center and an
envelope elongated in the south-east north-west direction. 
In Fig.~2 we show the velocity map of 
the HCO$^+$ 1$\rightarrow$0 line. A clear velocity gradient is detected in the south-east north-west direction. The direction 
of this gradient is close to orthogonal to the main Herbig Haro outflow, and consistent with the
existence of a rotating circumbinary disk.
The integrated emission of the HCN 1$\rightarrow$0 line is composed of a compact source and an extended envelope. The compact source peaks 
$\sim$1$"$ to the North of the phase center suggesting a larger contribution of GV~Tau~N to the total
emission. The extended emission is very asymmetric, being more extended towards the south-east than towards the north-west. Moreover, 
the HCN 1$\rightarrow$0 emission has an elongation in the outflow
direction (see Fig.~2). There is some hint of velocity gradient in the HCN 1$\rightarrow$0 
line although in this case the gradient is less clear than in the case of HCO$^+$~1$\rightarrow$0.

In Fig.~1 we show the spectra of the HCN 3$\rightarrow$2 and HCO$^+$ 3$\rightarrow$2 lines towards GV Tau N and S. The HCN 3$\rightarrow$2 line is
only detected towards the N component. The HCO$^+$ line is clearly detected towards the S component and tentatively detected towards the N.
The velocity integrated intensity map of the  HCN 3$\rightarrow$2 line is well fitted with an elliptical Gaussian of 0.32$\pm$0.04$"$$\times$0.25$\pm$0.05$"$ 
centered at the star position, a size slightly larger than that of the 1.1~mm continuum emission. The emission of the HCO$^+$ 3$\rightarrow$2  line is 
clearly more extended than the 1.1~mm continuum suggesting some contribution from the circumbinary disk and/or the nebula. Fitting a elliptical Gaussian in 
the uv plane, we obtain a size of 1.26$\pm$0.06$"$$\times$0.73$\pm$0.09$"$. 
The zero and first momentum maps of the HCN 3$\rightarrow$2 and HCO$^+$ 3$\rightarrow$2 line emissions are shown in Fig.~3. A clear velocity 
gradient is detected in GV~Tau~N. The velocity gradient defines a rotation axis that is 
in between that of the main Herbig Haro outflow and the Herbig jet (Devine et al. 1999, Movsessian \& Magakian 1999).  
In GV~Tau~S we detect a velocity gradient in the north-south direction but it
cannot be interpreted as a disk rotation. The kinematics of the HCO$^+$ 3$\rightarrow$2 line is very likely affected by the outflow motion.

\subsection{HCN and HCO$^+$ Column Densities}

Gibb et al. (2007, see also erratum 2008) detected absorption due to the HCN $\nu_3$ towards GV~Tau~N. The estimated column density and
rotational temperature were 3.7$\pm$0.3~10$^{16}$ cm$^{-2}$ and 115$\pm$10~K for HCN assuming a linewidth of 12~km~s$^{-1}$. With these physical 
conditions the HCN 3$\rightarrow$2 line is optically thick and the expected brightness temperature is $\sim$110 K.
We measured a peak intensity of 192~mJy/beam ($\sim$26~K) towards GV~Tau~N which implies that the size of the emitting region is $<$24~AU,
so that the warm HCN is present in the inner $R$$<$12~AU from the star. 
The HCN~3$\rightarrow$2/HCO$^+$ 3$\rightarrow$2 flux ratio towards GV~Tau~N is $\sim$10. LVG calculations
show that assuming that both molecules arise 
in the same region with the physical conditions derived by Gibb et al. (2007), the HCN/HCO$^+$ column density ratio is $\sim$300 (see Fig.~4a). 
The brightness temperature of the HCN~1$\rightarrow$0 line in this small region would then be $\sim$94~K, 
which diluted in the 3.4~mm beam would give a main beam brightness temperature of $<$0.28~K. The main beam brightness temperature we measure is 
larger, $\sim$0.85~K, consistent with the fact that the emission is extended and mostly 
arises from the outer part of the disk and the circumbinary disk/nebula.

Towards GV Tau S, only the HCO$^+$~3$\rightarrow$2 line is well detected and the HCN 3$\rightarrow$2/HCO$^+$ 3$\rightarrow$2 flux ratio 
is $<$0.3. Assuming n(H$_2$)=10$^7$~cm$^{-3}$ and a kinetic temperature of $\sim$18~K, we derive a beam averaged column density of 
N(HCO$^+$)=1.0$\times$10$^{14}$~cm$^{-2}$ in a linewidth of 8.6~km~s$^{-1}$ and the HCN/HCO$^+$ column density ratio would be $<$0.6 (see Fig.~4a). 
Other possibility is that
the emission arise in the lower density and warmer disk surface. In Fig.~4a, we explored this possibility (n(H$_2$)=10$^5$~cm$^{-3}$,
T$_k$=115~K) and conclude that in this case we would obtain N(HCN)/N(HCO$^+$)$<$1.6.
 
We use the 3.4~mm lines of HCN and HCO$^+$ to have an estimate of the HCN/HCO$^+$ column density ratio in the nebula surrounding 
the binary. Using the integrated line intensity of the weakest satellite component of the HCN 1$\rightarrow$0 line, assuming Local Thermodynamic
Equilibrium and T$_{rot}$=15~K,  we estimate a beam averaged column density of 1.7$\times$10$^{13}$~cm$^{-2}$. With the same assumptions, we
derive a HCO$^+$ column density of $\approx$1$\times$10$^{13}$~cm$^{-2}$. Therefore the average HCN/HCO$^+$ column density in this region is $\approx$1.7. 
This ratio is not dependent on the assumed rotation temperature as long as it is the same for the two molecules, which is a reasonable assumption
taking into account their similar dipole moments.

\begin{figure*}
\includegraphics{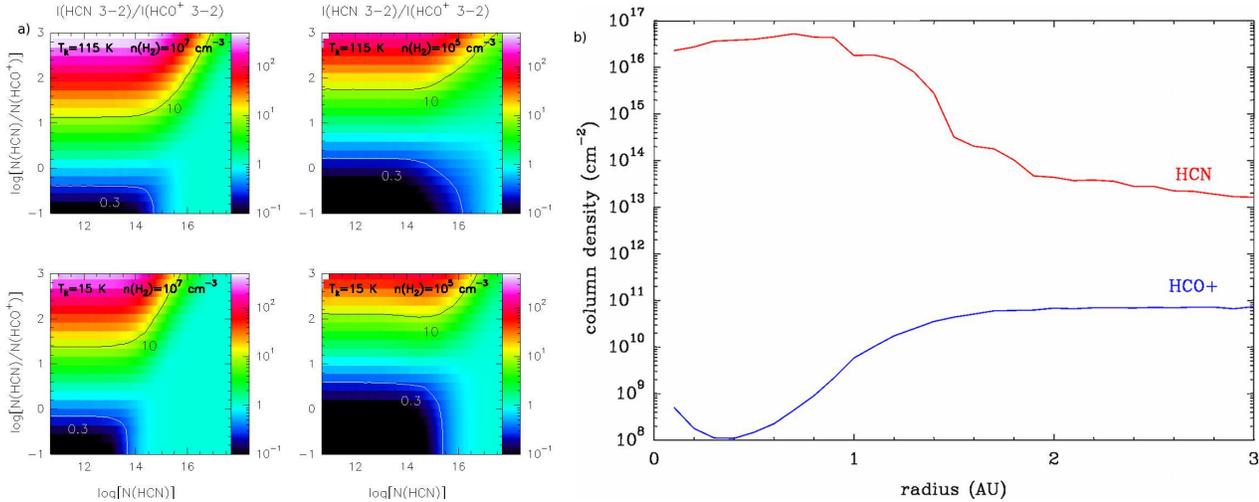}
\vspace{7.5cm}
      \caption{{\bf a)} Plots of the HCN 3$\rightarrow$2/HCO$^+$3$\rightarrow$2 line intensity ratio, I(HCN3$\rightarrow$2)/I(HCO$^+$3$\rightarrow$2)  
as a funcion of the HCN column density in a linewidth of 10~km~s$^{-1}$ and the HCN/HCO$^+$ abundance ratio for different physical conditions. The calculations have 
been done using the LVG code MADEX (J. Cernicharo 2012). Contour levels for the values found in GV~Tau~N, 10, and GV~Tau~S, 0.3, are drawn.
{\bf b)} HCN and HCO$^+$ vertical column densities calculated with
the model described in Ag\'undez, Cernicharo \& Goicoechea (2008) in the PDR within the inner 3 AU of a protoplanetary disk around a T Tauri star. 
They are computed by integrating the gas molecular
densities from the inner height $z_{in}$ (A$_v$$<$2) to the disk surface. 
}
         \label{Fig 4}
 \end{figure*}

\section{Discussion and Conclusions \label{sdis}}

Our millimeter data allow for the first time to sample the planet formation zone of the disks GV Tau N and GV Tau S and evidence
a dramatic chemical differentiation in the molecular gas associated with GV~Tau~N and GV~Tau~S. Previous NIR observations did not allow
to conclude because the absorption lines are strongly dependent on the disk orientation. Our observations show that the
HCN/HCO$^+$ ratio must be $>$300 in the planet formation zone of GV~Tau~N. In contrast, the HCN/HCO$^+$ must be $<$1.6 in GV~Tau~S.
The huge HCN column density found in the inner R$<$12~AU disk of GV~Tau~N is consistent with the chemical calculations by
Ag\'undez  Cernicharo \& Goicoechea (2008). They predicted an HCN abundance as
large as 5$\times$10$^{-5}$ in the inner R$<$3~AU of circumstellar disks (see Fig.~4b). The HCN is not arising from the disk mid-plane but from the 
photon-dominated region in the disk surface. The high densities and temperatures in this region are essential to achieve
a huge HCN abundance (Cernicharo 2004). 
The synthesis of organic molecules
in the gas requires that atomic carbon, produced by
the dissociation of CO, incorporates into C-bearing species
faster than reverting to CO.
For example, C$_2$H$_2$ and
HCN reach low abundances at 100 K because the reactions of C$_2$H and CN with H$_2$ have activation barriers
(Cernicharo 2004).
At temperatures
above $\sim$400~K atomic oxygen is efficiently converted
into OH, which may react with C to form CO but reacts
faster with H$_2$ to form water. Thus, most of the oxygen
forms H$_2$O, and CO does not reach its maximum abundance
allowing atomic carbon to form C-bearing molecules. 
The key reactions for the HCN formation
such that C+NO$\rightarrow$CN+O, or H$_2$+CN$\rightarrow$HCN+H,
proceed very rapidly at these large gas densities ($>$10$^7$~cm$^{-3}$) and temperatures (Cernicharo 2004).
This mechanism also works for any dense PDR such as those found around protoplanetary nebula where the
photodissociation of CO and HCN allows a fast photo-polymerization towards longer carbon chains (Cernicharo 2004).This same
result was found by Walsh et al. (2010) in their disk chemical model which included a grain-surface network. In 
that model the [HCN]/[HCO$^+$] ratio was $\sim$1000 at R$\sim$1~AU and always $>$10 for R$<$10~AU.
Although not essential, accretion shocks could contribute to enhance the density and gas kinetic temperature and make the HCN formation 
more efficient.

One puzzling question then is the non-detection of HCN in GV~Tau~S. This can only be explained by 
a different disk morphology (flatter disk, an inner gap), different dust properties and/or a different gas/dust ratio
in the inner disk. Higher spatial resolution interferometric observations are required to unveil the
hidden structure of this interesting binary.


\acknowledgments
\begin{small}
This project has been partially supported within the program CONSOLIDER INGENIO
2010, under grant  CSD2009$-$00038 $"$Molecular Astrophysics: The Herschel and ALMA Era$–$
ASTROMOL$"$. We also thank the Spanish MICINN for funding support through
grants AYA2006-14876 and AYA2009-07304. 
\end{small}



\end{document}